\documentclass[12pt]{article}
\usepackage{amsmath,amssymb,epsfig}
\usepackage{graphicx}
\usepackage{makeidx}
\usepackage{cite}

\def\CPPP{\mathbb{C}\mathrm{P}^3}
\def\CPP{\mathbb{C}\mathrm{P}^2}
\def\CP{\mathbb{C}\mathrm{P}^1}
\def\SYMNNnk{(\mathrm{Sym}({\bf N}^{nk}), \mathrm{Sym}({\overline{\bf N}}^{nk}))}
\def\SYMNNk{(\mathrm{Sym}({\bf N}^{k}), \mathrm{Sym}({\overline{\bf N}}^{k}))}
\def\CZ{(\Bbb{C}^4/\Bbb{Z}_k)^N}

\def \Tr{{\rm{Tr}}}

\def \be{\begin{equation}}
\def \ee{\end{equation}}
\def \beq{\begin{eqnarray}}
\def \eeq{\end{eqnarray}}

\makeatletter

\renewcommand\section{\@startsection {section}{1}{\z@}%
                                   {-3.5ex \@plus -1ex \@minus -.2ex}%
                                   {2.3ex \@plus.2ex}%
                                   {\normalfont\large\bfseries}}

\renewcommand\subsection{\@startsection{subsection}{2}{\z@}%
                                     {-3.25ex\@plus -1ex \@minus -.2ex}%
                                     {1.5ex \@plus .2ex}%
                                     {\normalfont\normalsize\bfseries}}

\newcount\hour \newcount\minute
\hour=\time \divide \hour by 60
\minute=\time
\def\now{%
\ifnum \hour<13
  \ifnum \hour=0 \advance \hour by 12 \number\hour:\else \number\hour:\fi%
     \ifnum \minute<10 0\fi%
     \number\minute%
\ A.M.%
\else \advance \hour by -12 \number\hour:%
  \ifnum \minute<10 0\fi%
  \number\minute%
  \ P.M.%
\fi%
}

\makeatother

\begin{document}

\baselineskip=18pt  
\numberwithin{equation}{section}  
\allowdisplaybreaks  



%
%


\thispagestyle{empty}

\vspace*{-2cm}
\begin{flushright}
{\tt arXiv:0810.1075}\\
CALT-68-2699\\
\end{flushright}


\vspace*{1.7cm}
\begin{center}
 {\large {\bf Comments on Baryon-like Operators}\\
 {\bf in $\mathcal{N}=6$ Chern-Simons-matter theory of ABJM}}

 \vspace*{2cm}
 Chang-Soon Park\\
 \vspace*{1.0cm}

{\it California Institute of Technology 452-48, Pasadena, CA 91125, USA}\\
 \vspace*{0.8cm}

{\tt p{}a{}rk{}@c{}a{}l{{}}te{{}}ch.e{{}}d{}u} 

\end{center}
\vspace*{.5cm}


We show that baryon-like operators exist in the $\mathcal{N}=6$ superconformal Chern-Simons-matter theory constructed by Aharony, Bergman, Jafferis and Maldacena (ABJM).
This involves the introduction of Wilson(or 't Hooft) lines ending on baryon-like operators and we show that the presence of such lines cannot be detected by the fields in the theory.
The same construction can be used to make magnetic monopoles charged under the diagonal $U(1)$ gauge group with arbitrary non-integral charge.
If we do not include such monopole configurations in the path integral, the moduli space is known to be given by the symmetric $N$ copies of $\Bbb{C}^4/\Bbb{Z}_k$.
However, if we allow for such monopole configurations, the flux quantization conditions change so that we do not have a discrete gauge symmetry and effectively the $(SU(N)\times SU(N))/\Bbb{Z}_N$ gauge theory remains.
We discuss the possibility of the level-rank duality of the theory.


\newpage
\setcounter{page}{1} 





\section{Introduction}
In this paper, we are going to discuss baryon-like chiral operators in the superconformal Chern-Simons-matter theory constructed in \cite{Aharony:2008ug}(ABJM). This work has obtained greater interest because it can be used to describe multiple M2-branes.
The possibility for constructing the Lagrangian description for multiple M2-branes with Chern-Simons-matter theories was explored in \cite{Schwarz:2004yj}.
A concrete description for a pair of M2-branes has been obtained in \cite{Bagger:2006sk,Bagger:2007jr,Bagger:2007vi,Gustavsson:2007vu}.
ABJM generalize the theory to an arbitrary number of M2-branes.
In their description, the level $k$ of the Chern-Simons action is related to the orbifolding of the transverse space $\Bbb{C}^4/\Bbb{Z}_k$.
For general $k$, the orbifold procedure leaves only $\mathcal{N}=6$ supersymmetry and we cannot see $\mathcal{N}=8$ for level $k=1,2$ manifestly.
Various aspects of this theory were explored in subsequent papers and a partial list of them is \cite{Krishnan:2008zm,Hosomichi:2008jd,Bandres:2008vf,Papadopoulos:2008sk,Gauntlett:2008uf,Bandres:2008kj,Gomis:2008be,Benvenuti:2008bt,Benna:2008zy,Bhattacharya:2008bja,Nishioka:2008gz,Minahan:2008hf,Gaiotto:2008cg,Ahn:2008gda,Grignani:2008is,Hosomichi:2008jb,Bagger:2008se,Bak:2008cp,McLoughlin:2008ms,Alday:2008ut,Krishnan:2008zs,Ooguri:2008dk,Jafferis:2008qz,Berenstein:2008dc,Hosomichi:2008ip,Ahn:2008ua,Klebanov:2008vq,McLoughlin:2008he}.

ABJM presented various chiral operators. Since we have a $U(1)$ factor in the $U(N)\times U(N)$ gauge group that couples to the matter fields, naively an operator with nonzero $U(1)$ charge is not allowed. However, they showed that an operator of the schematic form $C^{nk}$ in the $nk$'th symmetric product of the $\mathbf{4}$ representation of the $SU(4)_R$ R-symmetry, where $C$ denotes the bosonic matter fields, can be attached to the end of a Wilson(or 't Hooft) line so that the resulting operator becomes gauge-invariant.
Usually, the introduction of a Wilson line can be detected by the fields in the theory and makes the operator non-local.
But in this special case with the Chern-Simons action, any fields in the theory cannot detect the Wilson line and the chiral operator $C^{nk}$ with the Wilson line is a good local operator.

In the $SU(N)\times SU(N)$ theory, we can also make a baryon-like operator whose schematic form is $\det (C)$.
In the $U(N)\times U(N)$ theory, this operator is not gauge-invariant.
To make it gauge-invariant, we attach a Wilson line with suitable $U(1)$ charge to the operator analogously to what happens to the operator $C^{nk}$. We will show that the Wilson line cannot be detected by any fields in the theory. This is possible because of the special assignment of levels ($k$ and $-k$).
The argument actually shows that we can place a magnetic monopole charged under the diagonal $U(1)$ with non-integral charge.
If we choose to include such configurations in the path integral, the flux quantization conditions change and we effectively have the $(SU(N)\times SU(N))/\Bbb{Z}_N$ theory without discrete gauge symmetry in the conformal phase in which the gauge group in unbroken.
In this case, the moduli space is different from the symmetric $N$ copies of $\Bbb{C}^4/\Bbb{Z}_k$.
Therefore, we have to consider $U(N)\times U(N)$ gauge theory with only integral magnetic charges to obtain the correct moduli space for $N$ M2-branes.
However, to avoid confusion, let us here stress that the baryon-like operator we analyzed is allowed even in the theory with only integral magnetic charges.
That is, the existence of such baryon-like operators does not depend on which flux quantization condition we choose.

The paper is organized as follows. In Section \ref{S:BaryonOp}, we review the known chiral operators in the $\mathcal{N}=6$ superconformal Chern-Simons-matter theory of ABJM and construct baryon-like operators by attaching appropriate Wilson lines. In Section \ref{S:FluxQuant}, we consider the flux quantization conditions and show that we have the $(SU(N)\times SU(N))/\Bbb{Z}_N$ gauge theory without discrete gauge symmetry if we allow for non-integral diagonal monopoles.
In Section \ref{S:LevelRank}, we discuss the result and speculate about a duality of the theory by which the level $k$ and the rank $N$ are exchanged.
In Appendix \ref{A:SUGRA}, we calculate the scaling dimensions and the $SU(4)_R$ representations of the operators $C^k$ and $\det(C)$ in the supergravity side and confirm they agree with the field theory expectation.

\section{Baryon-like chiral operators}\label{S:BaryonOp}
In this section, we will show how baryon-like operators arise in the $\mathcal{N}=6$ superconformal Chern-Simons-matter theory of ABJM.
But first, let us list various chiral operators found in \cite{Aharony:2008ug}.

\subsection{Chiral operators}\label{SS:ChiralOp}
The theory has four bosonic matter fields $C_I$ in the $\mathbf{4}$ representation of the $SU(4)_R$ R-symmetry.
They are also in the $(\mathbf{N}, \overline{\mathbf{N}})$ representation of the $U(N)\times U(N)$ gauge group.
There are two gauge fields $A_{(1)}$ and $A_{(2)}$ and the fields $C_I$ have the covariant derivative
\begin{equation}
D_{\mu} C_I = \partial_{\mu} C_I + i(A_{(1)\mu} C_I - C_I A_{(2)\mu})\;.
\end{equation}
Especially, the $U(1)$ factor in each gauge group couples to $C_I$ via the term $i(a_{(1)\mu} - a_{(2)\mu}) C_I$ where $a_{(i)}=\frac 1 N \Tr A_{(i)}$ is the trace part of the gauge field $A_{(i)}$.
Let $U(1)_{\tilde{b}}$ be the diagonal $U(1)$ subgroup of $U(N)\times U(N)$ and $U(1)_{b}$ the anti-diagonal one. That is, $U(1)_{\tilde{b}}$ is associated to the gauge field $a_{\tilde{b}}=a_{(1)}+a_{(2)}$ and $U(1)_b$ to $a_b=a_{(1)}-a_{(2)}$.
Note that the matter fields are charged only under $a_b$ and not $a_{\tilde{b}}$.

One class of operators has the schematic form $\Tr((C_I C^{\dagger}_J)^l)$. They are in the $(l,0,l)$ representation of $SU(4)_R$ using the Dynkin labels.
In other words, they are in the $l$'th symmetric product of $\mathbf{4}$ and $\bar{\mathbf{4}}$, respectively, with trace parts subtracted. They do not carry $U(1)_b$ charge.

Another class of chiral operators is written schematically as $C^{nk}$(or $C^{\dagger nk}$). They are in the $nk$'th symmetric product of the $\mathbf{4}$ representation, or in $(nk,0,0)$ of $SU(4)_R$.
An operator in this class in itself is not gauge-invariant, but we can attach a Wilson line to it.
That is, we put one end of a Wilson line to the operator and the other end goes to infinity.
The relevant Wilson line is in the $\SYMNNnk$ representation.
In three dimensions, a Wilson line in some representation is equivalent to an 't Hooft line \cite{Moore:1989yh} and, for this $\SYMNNnk$ representation, the corresponding 't Hooft line is not observable by any fields in the theory \cite{Itzhaki:2002rc}.

In addition to this, we can also consider a baryon-like chiral operator of the form $\det(C)$.
It carries $N$ units of $U(1)_b$ charge and to compensate for this, we need to attach a Wilson line with $U(1)_b$ charge to it.
Below we will review the attachment of Wilson lines in more detail and see that a Wilson line attached to a baryon-like operator cannot be detected by any fields in the theory.

\subsection{Wilson lines in three dimensions}\label{SS:WilsonLines}
Before considering the product gauge group $U(N)\times U(N)$, let us review the relation between a Wilson loop and an 't Hooft loop in three dimensions in the Abelian and $SU(N)$ gauge groups.

In three dimensions, an 't Hooft loop is equivalent to a Wilson loop in the presence of the Chern-Simons action\cite{Moore:1989yh}\footnote{There is a Wilson loop that does not correspond to any 't Hooft loop.}.
As a simple example, let us consider the Abelian $U(1)$ Chern-Simons theory.
It is given by
\begin{equation}\label{E:AbelianCSS}
S=\frac k {4\pi} \int a \wedge da\;.
\end{equation}


\begin{figure}[t]
\begin{center}
  \epsfxsize=5.0cm \epsfbox{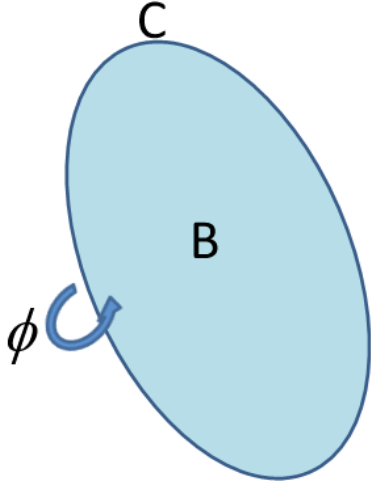}
\caption{\sl
 't Hooft loop $C$}
\label{Figure1}
\end{center}
\end{figure}

Consider a closed loop $C$ and perform a gauge transformation $\Lambda(\phi)=\alpha \phi$ near the curve $C$. $\alpha$ is a constant and $\phi$ is the angular variable winding the loop. It creates a magnetic flux tube along the loop $C$ with total magnetic flux $2\pi \alpha$. Call such a procedure $T_\alpha[C]$. It is an 't Hooft loop \cite{'tHooft:1977hy}. Let $B$ be a (any) surface bounded by $C$(See Figure \ref{Figure1}). The angular variable has value 0(or $2\pi$) on the surface $B$.
Such gauge transformation $a\rightarrow  a+d\Lambda$ changes the action \eqref{E:AbelianCSS}:
\begin{equation}
\begin{split}
\delta S &= \frac k {2\pi} \int d(\Lambda \wedge da)\\
&=\frac k {2\pi} \left(\int_{\text{right side of B}} - \int_{\text{left side of B}}\right)(\Lambda \wedge da)\\
&=k \alpha \int_B da = k \alpha \int_C a\;.
\end{split}
\end{equation}
So the insertion of this operator $T_\alpha[C]$ is equivalent to the insertion of the Wilson loop $W_\alpha [C]$
\begin{equation}
W_\alpha[C]=e^{i k \alpha \int_C a}
\end{equation}
to the action.

\begin{figure}[t]
\begin{center}
  \epsfxsize=6.0cm \epsfbox{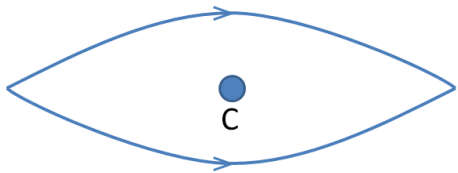}
\caption{\sl
 Interference of a particle due to the magnetic flux through the curve $C$}
\label{Figure2}
\end{center}
\end{figure}

How can we detect the presence of the operator $T_\alpha[C]$(or equivalently $W_\alpha[C]$)? One way is to measure the interference due to the Aharonov-Bohm effect. Consider two paths of a particle of charge $q$ under $U(1)$ above and below the curve $C$ as shown in Figure \ref{Figure2}.


The difference of the phase of the wave function due to the magnetic flux through the curve $C$ is $q \oint A= q \oint d\Lambda = 2\pi q\alpha$.
That means the following. If we search the 't Hooft operator $T_\alpha [C]$ using a particle of charge $q=+1$, we cannot detect $T_\alpha[C]$ if $\alpha$ is an integer. But if we use a particle of charge $q=+2$ instead, $T_\alpha[C]$ with half integral $\alpha$ will still not be detected.
Put differently, whenever the gauge transformation $\Lambda(\phi)$ at $\phi=2\pi$ leaves invariant the particle we use to detect the loop, $T_\alpha[C]$ is unobservable.

The analysis can be extended to non-Abelian gauge theories also, as was done in \cite{Moore:1989yh,Itzhaki:2002rc}.
The final result is as follows. Consider the $SU(N)$ gauge theory in which all fields are invariant under the center of $SU(N)$.
Let $T[C]$ be a prescription related to a curve $C$ such that the gauge transformation around the curve $C$ is given by
\begin{equation}\label{E:gaugetrans}
g(\phi)= \exp (i \phi H)\;,
\end{equation}
where $\phi$ is the angular variable around the curve $C$ and $H=\mathrm{diag}(\frac 1 N,\frac 1 N,\cdots,\frac 1 N,\frac 1 N - 1)$.\footnote{To be precise, we have to average over all possible gauge transformations of \eqref{E:gaugetrans} to define a gauge-invariant operator since \eqref{E:gaugetrans} chooses a preferred direction determined by $H$.}
Then it can be shown that for $k=1$, $T[C]=W_{\bf N}[C]$, where $W_{\mathbf{N}}[C]$ is the Wilson-loop in the fundamental $\mathbf{N}$ representation.
For $k>1$, $T[C]=W_{\mathrm{Sym}({\bf N}^k)} [C]$, where $W_{\mathrm{Sym}({\bf N}^k)} [C]$ is the Wilson-loop in the $k$'th symmetric product of the $\mathbf{N}$ representation.

It looks as if $T[C]$ is a local operator since all fields in the theory are assumed to be invariant under the center of the gauge group.
However, since $T[C]$ is equivalent to $W_{\mathrm{Sym}({\bf N}^k)} [C]$, which is sensitive to the center, an 't Hooft loop can be detected by another 't Hooft loop. So the proper interpretation is that an 't Hooft line(not a loop) with one end at some point and the other at infinity defines an anyon at that point \cite{Itzhaki:2002rc}.
This will be different from the theory with the product gauge group we are interested in, as will be shown in the next section.

\subsection{Baryon-like operators with Wilson lines}\label{SS:Baryons}
Here we will consider an operator of the form $\det(C)$ in the ABJM theory. More precisely, this operator has the form
\begin{equation}
\epsilon_{i_1 \cdots i_N} \epsilon^{\hat j_1\cdots \hat j_N} C_{I_1 , \hat j_1}^{i_1}\cdots C_{I_N, \hat j_N}^{i_N}\;,
\end{equation}
where $I_a$ is for $\mathbf{4}$ of $SU(4)_R$, $i_a$ for $\mathbf{N}$ of one $U(N)$ and $\hat j_a$ for $\overline{\mathbf{N}}$ of another $U(N)$. Note that the two epsilon tensors ensure that all flavor indices $I_a$ are symmetrized.
This operator carries $N$ units of $U(1)_b$ charge.
To compensate for this, we have to attach the following Wilson line to the operator at the point $x$.
\begin{equation}\label{E:AbelWilson}
W[C]=e^{i N \int_x^{\infty} a_b}=e^{i N \int_x^{\infty}(a_{(1)}-a_{(2)})}\;,
\end{equation}
where $a_b=a_{(1)}-a_{(2)}$ is the Abelian part of the gauge fields, as defined in Section \ref{SS:ChiralOp}.
Extending the argument of Section \ref{SS:WilsonLines} to the product gauge group, this Wilson line is equivalent to an 't Hooft line around which the gauge transformation is given by\footnote{Unlike the Abelian example in Section \ref{SS:WilsonLines}, the level for the Abelian part is $N k$ since the Abelian part of the gauge field is $a_{(i)}=\frac 1 N \Tr A_{(i)}$.}
\begin{equation}
(e^{\frac i k \phi},e^{\frac i k \phi})\in U(N)\times U(N)\;.
\end{equation}
Note that the matter fields $C_I$(and their fermionic partners) are insensitive to the presence of the Wilson line since they are invariant under the gauge transformation $(e^{\frac {2\pi i} k},e^{\frac {2\pi i} k})\in U(N)\times U(N)$.
Actually, the same argument can be used to say that we can have a Wilson line with arbitrary $U(1)_b$ charge at the end of it and the line still cannot be detected by the matter fields. This Wilson line is equivalent to an 't Hooft line, which in turn can be thought of as a monopole\footnote{We call it a monopole despite the fact that it is localized in both space and time in three dimensions. Note that the signature of the spacetime is not important in this case since the monopole configuration only involves the gauge fields, which does not have kinetic terms.} magnetically charged under $U(1)_{\tilde{b}}$ \cite{Borokhov:2002ib,Borokhov:2002cg,Kapustin:2005py}.

We can also directly show that the Wilson line \eqref{E:AbelWilson} defines a local operator.
It will be local if the integration from $x$ to $\infty$ in \eqref{E:AbelWilson} does not depend on the path.
Then it is enough to check
\begin{equation}
e^{i N \oint a_b}=1\;,
\end{equation}
for any closed path.
But the Abelian part of the Chern-Simons action is proportional to $\int da_b \wedge a_{\tilde{b}}$ and this is the only place where $a_{\tilde{b}}$ appears.
Therefore, the equation of motion for $a_{\tilde{b}}$ sets $da_b$ to zero and this implies $\oint a_b=0$ for any closed path\footnote{Of course, $da_b=0$ does not directly imply $\oint a_b=0$. This will not be true if we have a monopole field configuration with magnetic $U(1)_b$ charge. This is equivalent to a Wilson line in a representation of $U(1)_{\tilde{b}}$. However, in such a configuration, there is a magnetic flux coming out of the monopole so that $da_b=0$ cannot be satisfied. Therefore, the only configurations that contribute to the path integral are those without $U(1)_b$ monopoles. In that case, $da_b=0$ does imply $\oint a_b=0$. Instead of monopoles, there can be an infinitely extended vortex line charged under $U(1)_b$. That is, around the vortex line we perform a gauge transformation $(e^{i m \phi},e^{-i m \phi})\in U(N)\times U(N)$. This is equivalent to an infinite Wilson line in a representation of $U(1)_{\tilde{b}}$. If $m$ is an integer, integrating $a_b$ around the line gives $\oint a_b=2\pi m$. Hence $e^{N\oint a_b}=1$, and again the expectation value of the Wilson line does not change. Of course, we can have a vortex line with non-integral $m$ and, in that case, our Wilson line can give a different expectation value as it crosses the vortex line. However, such a vortex line is not a field configuration that we include in the path integral and it is fine to have two non-local external objects in three dimensional spacetime just as the fractional statistics of anyons. The same conclusion can be made for an infinitely extended(or circular) Wilson loop with $U(1)_{\tilde{b}}$ charge.}.
Therefore we again conclude that \eqref{E:AbelWilson} defines a local operator.

\begin{figure}[t]
\begin{center}
  \epsfxsize=5.0cm \epsfbox{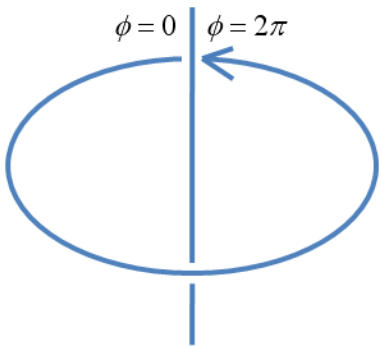}
\caption{\sl
 't Hooft loop winding around another 't Hooft loop}
\label{Figure3}
\end{center}
\end{figure}

As stated at the end of Section \ref{SS:WilsonLines}, an 't Hooft loop winding around another 't Hooft loop obtains a non-trivial phase in the $SU(N)$(or $U(N)$) gauge theory.
However, in the $U(N)\times U(N)$ theory considered here, the operators defined by the end points of such 't Hooft lines do not see each other's 't Hooft line.
To see this, let's wind an 't Hooft loop with gauge transformation $(g(\phi), g(\phi))$ around another 't Hooft loop with the same gauge transformation but extending as a straight line(see Figure \ref{Figure3}). $g(\phi)$ for each $U(N)$ is as defined in \eqref{E:gaugetrans}.
Treating the winding 't Hooft loop as a Wilson loop in the $\SYMNNk$ representation, we see that the Wilson loop will pick up a phase $(g(2\pi)^k, g(2\pi)^{k})=(e^{2\pi k i /N}, e^{2\pi k i/N}) \in U(N)\times U(N)$. That is, the Wilson loop $W[C]$ becomes $g(2\pi)^k W[C] g(2\pi)^{-k}=W[C]$.
It is invariant.
In the same way, any Wilson line attached to any operator we considered so far does not give a nontrivial phase to a Wilson loop surrounding the line.
Therefore, we conclude that all chiral operators considered in \cite{Aharony:2008ug} can be defined in such a way that the Wilson lines(or 't Hooft lines) attached to them cannot be detected by other such operators, nor by the matter fields.


\section{Flux quantization conditions}\label{S:FluxQuant}
In the previous section, we see that we can have a monopole charged under $U(1)_{\tilde{b}}$ with arbitrary charge\footnote{But the baryon-like operators are allowed even when only integrally-charged monopoles are allowed.}.
In the computation of the moduli space, we can either include such monopole configurations or not in the path integral.
This amounts to the choice of the gauge group.
If we choose the gauge group to be $U(N)\times U(N)$ without any identification of the center of the group, monopoles charged under $U(1)_{\tilde{b}}$ can have only integral charges.
However, if we identify elements related by the diagonal $U(1)_{\tilde{b}}$ in the gauge group, such monopoles are allowed.
In this case, the more precise gauge group is $(U(N)\times U(N))/U(1)_{\tilde{b}}$.
Below, we will consider the theory with this gauge group.

First, let us focus on the conformal phase in which the gauge group is unbroken.
The Chern-Simons action has the Abelian parts
\begin{equation}
S_{CS}=\frac {N k} {4\pi} \int \left(a_{(1)} \wedge da_{(1)} - a_{(2)} \wedge da_{(2)}\right)=\frac {N k} {4\pi} \int a_b \wedge da_{\tilde{b}}\;.
\end{equation}
Since this is the only place where $f_{\tilde{b}}=da_{\tilde{b}}$ appears, we can introduce a dual variable $\tau(x)$ and add to the action\cite{Distler:2008mk,Lambert:2008et,Aharony:2008ug}
\begin{equation}
S_{\tau} = \frac N {4\pi} \int \tau(x) \epsilon^{\mu\nu\lambda} \partial_{\mu} f_{\tilde{b}\nu\lambda}\;.
\end{equation}
The equation of motion of $f_{\tilde{b}\nu\lambda}$ gives $a_{b\mu} = \frac 1 k \partial_{\mu} \tau$.
A Wilson line carrying $q$ units of $U(1)_b$ charge is equivalent to an 't Hooft line which gives the gauge transformation $(e^{i \frac q {N k} \phi},e^{i \frac q {N k} \phi})\in U(N)\times U(N)$ around the line. Note that this gauge transformation is related to the $U(1)_{\tilde{b}}$ group. Since $q$ can have an arbitrary value, the magnetic flux of $U(1)_{\tilde{b}}$ that comes out of the end of the 't Hooft line can be arbitrary. That is, $\int f_{\tilde{b}}$ is not restricted to an integer value. Therefore, once we gauge-fix $\tau$ to 0, there is no remaining discrete gauge symmetry.
In other words, we have effectively the $(SU(N)\times SU(N))/\Bbb{Z}_N$ gauge theory.

Now let us consider the moduli space in the Coulomb phase\footnote{We assume that, as in \cite{Aharony:2008ug}, the classical moduli space is not modified by the quantum corrections.} taking into account the diagonal monopoles with non-integral magnetic charge.
When the gauge group is broken to $U(1)_b^N \times U(1)_{\tilde{b}}^N$, for each $F_{\tilde{b} i}=dA_{\tilde{b} i}$ where $i=1,\cdots, N$, we introduce a dual variable $\tau_i(x)$ and add a Lagrange multiplier to the original action and make $F_{\tilde{b} i}$ instead of $A_{\tilde{b} i}$ as  the basic variable.
Then, we have in the action
\begin{equation}\label{E:LagMul}
S=\cdots+\frac k {4\pi} \sum_{i=1}^N \int A_{b i} \wedge F_{\tilde{b} i} + \frac 1 {4\pi} \sum_{i=1}^N \int \tau_i (x) \epsilon^{\mu\nu\lambda} \partial_{\mu} F_{\tilde{b} i \nu \lambda}+\cdots\;.
\end{equation}
In the Coulomb phase, we can construct many kinds of monopoles.
Each monopole can be uniquely determined by its gauge transformation around the Dirac string(or 't Hooft line) to which it is attached.
One type of monopole has the gauge transformation
\begin{equation}\label{E:monopoleGT}
(e^{i \phi T}, e^{i \phi T})\in U(N)\times U(N)\;,
\end{equation}
around the Dirac string and $T=\mathrm{diag}(1,0,\cdots,0)$.
The condition to be a good monopole is that the gauge transformation at $\phi=2\pi$ leaves the fields in the theory invariant.
Obviously, this condition is met for this type of monopole.
We can permute the diagonal elements of $T$ and get additional $N-1$ types of monopoles.
Also, we have a diagonal monopole whose gauge transformation around the Dirac string is still given by \eqref{E:monopoleGT} but now $T=\mathrm{diag}(\alpha,\cdots,\alpha)$ for an arbitrary real number $\alpha$.

In the presence of such types of monopoles, the flux $\Phi_i$ for the $i$'th diagonal $U(1)_{\tilde{b}}$ through a sphere surrounding monopoles is given by
\begin{equation}
\Phi_i=2\pi(n_i + \alpha)\;,
\end{equation}
where each $n_i$ is an integer and $\alpha$ is the same for all $i$.
The action is defined up to the phase $2\pi$.
Therefore the set of functions $\tau_i(x)$ and the set of functions $\tau_i(x)+ \Delta_i$, where $\Delta_i$ are constants, will be identified if
\begin{equation}
\begin{split}
&\sum_i \Delta_i (n_i + \alpha)\in 2\pi \Bbb{Z}\\
\mathrm{or}\qquad &\sum_i \Delta_i n_i + \alpha \sum_i \Delta_i \in 2\pi \Bbb{Z}\;,
\end{split}
\end{equation}
for any set of integers $n_i$ and a real number $\alpha$.
Since $\alpha$ is real, we have $\sum_i \Delta_i=0$.
The first summation indicates that $\Delta_i \in 2\pi \Bbb{Z}$.
The equation of motion of $F_{\tilde{b} i}$ in \eqref{E:LagMul} says $A_{b i \mu}=\frac 1 k \partial_{\mu} \tau_i$.
Then the covariant derivative of the $i$'th diagonal element $C_{i I}$ of the matter field in the Coulomb phase becomes $D_{\mu} C_{i I} = \partial_{\mu} C_{i I}+ \frac i k C_{i I} \partial_{\mu} \tau_i$.
The gauge transformation $\tau_i \rightarrow \tau_i +\Delta_i$ changes $C_{i I}$ to $e^{-\frac i k \Delta_i} C_{i I}$.
Therefore, once we gauge-fix the variables $\tau_i$ to vanish, there still remain discrete gauge transformations $e^{-2\pi \frac i k n_i} C_{i I}\,$ if $\, \sum_i n_i =0$.
Call the resulting space $\mathcal{M}$.\footnote{We implicitly assume the identification by permutation of elements henceforth.}
Compared to the space $\CZ$, where $k^N$ points in $\Bbb{C}^{4N}$ are identified to a point, we have only $k^{N-1}$ points identified to a point in the space $\mathcal{M}$.
That is, a $k$-fold identification is missing compared to $\CZ$.
More specifically, given a point $(p_1,\cdots p_N)$ in $\Bbb{C}^{4N}$, let us define a function from a set of integers $(n_1,\cdots, n_N)$ to a point in $\Bbb{C}^{4N}$:
\begin{equation}\label{E:orbiaction}
P(n_1,\cdots,n_N)=(e^{2\pi \frac i k n_1} p_1, \cdots, e^{2\pi \frac i k n_N} p_N)\;,
\end{equation}
where, for example, $e^{2\pi \frac i k n_1} p_1$ means multiplying all four components of $p_1$ by the same factor $e^{2\pi \frac i k n_1}$.
Then $\CZ$ is defined by identifying all $P(n_1,\cdots, n_N)$, whereas in the space $\mathcal{M}$, the following $k$ points are different:
\begin{equation}
P(0,0,\cdots,0),\, P(1,0,\cdots,0),\,\cdots\, ,P(k-1,0,\cdots,0,0)\;.
\end{equation}
Note that $P(k,0,\cdots,0)$ is the same point as $P(0,\cdots,0)$ in $\Bbb{C}^{4N}$, so no gauge transformation is needed to identify them.
These $k$ points represent the $k$ equivalence classes.
Other points belong to one of the $k$ equivalence classes and the equivalence class for $P(n_1,\cdots, n_N)$ is determined by $\sum_i n_i$ mod $k$.

Note that the space $\mathcal{M}$ is a $k$-fold covering space of $\CZ$.
That we end up with the space $\mathcal{M}$ and not $\CZ$ hinges on the existence of the non-integral diagonal monopole, which is given by \eqref{E:monopoleGT} with $T=\mathrm{diag}(\alpha, \cdots, \alpha)$ and an arbitrary real number $\alpha$.
But we do not have to include such non-integral monopoles in the theory.
When we perform the path integral to calculate correlation functions of the theory, it is perfectly legitimate to restrict to configurations of the gauge fields corresponding only to integral monopoles.
More precisely, we can only consider gauge field configurations with monopoles that are consistent with fields in the fundamental representation of each gauge group.
That is, the gauge group is $U(N)\times U(N)$ without identification of the center.
In this case, the moduli space will turn out to be $\CZ$.
Since this is the correct moduli space for $N$ M2-branes, we conclude that we have to include only monopole configurations compatible with fields in the fundamental representation for each gauge group.
The Wilson line in \eqref{E:AbelWilson} is still well-defined in this setup and therefore the operator of the form $\det(C)$ can exist by adding this Wilson line.

\section{Summary and discussions about level-rank duality}\label{S:LevelRank}
In this paper, we show that baryon-like operators exist in the $\mathcal{N}=6$ superconformal Chern-Simons-matter theory of ABJM by adding Wilson lines with appropriate $U(1)_b$ charge. The Wilson lines are not observable by any fields or operators in the theory.
If we consider only a Wilson line with $U(1)_b$ charge from a point extending to infinity, the end of the Wilson line describes a monopole magnetically charged under $U(1)_{\tilde{b}}$.

The magnetic flux out of the monopole can have an arbitrary value without harming the consistency of the theory since no fields in the theory can detect the Dirac string attached to this monopole.
However, if we allowed for such monopole configurations in the path integral, the moduli space in the Coulomb phase would turn out to be a $k$-fold covering space of $\CZ$ and not $\CZ$ itself.
Therefore, such configurations would not be allowed in the Coulomb phase.
But, we may still think of the possibility of including such configurations in the conformal phase of the theory.
Although we do not know how this can be achieved exactly, we would like to point out some interesting consequence assuming this possibility.

Suppose we allow for non-integral diagonal monopole configurations in the conformal phase.
More precisely, we include in the path integral all possible gauge field configurations with monopoles that are consistent with fields in the theory: bi-fundamentals and adjoints.
In this case, the gauge group is effectively $(SU(N)\times SU(N))/\Bbb{Z}_N$ as shown in Section \ref{S:FluxQuant} since the $U(1)_b$ factor of the gauge group does not give a discrete gauge symmetry.
For example, for the $U(1)\times U(1)$ gauge theory, $U(1)_{\tilde{b}}$ does not couple to the matter fields from the beginning and $U(1)_b$ couples, but once the gauge field $A_b$ has been set to vanish, it does not give any additional discrete gauge symmetry.
Therefore we have a free theory with four matter fields without gauge fields.
In this case, we have $\mathcal{N}=8$ supersymmetry for any $k$\cite{Schwarz:2004yj,Bandres:2008ry}.
For the $U(2)\times U(2)$ gauge theory, in the same way, we effectively have the $(SU(2)\times SU(2))/\Bbb{Z}_2$ gauge theory in the conformal phase and we recover $\mathcal{N}=8$ supersymmetry for any $k$\cite{Bagger:2006sk,Bagger:2007jr,Bagger:2007vi}.
This looks interesting because we know that we have enhanced $\mathcal{N}=8$ supersymmetry for any $U(N)\times U(N)$ theory when $k=1$ or $2$.
So we suspect that there is some version of level-rank duality in the theory\cite{Naculich:1990hg}.

However, there is a subtle point that requires further examination.
Note that the $U(N)\times U(N)$ theory with level $k=1$ has four gauge-invariant operators $C_I$(attached to Wilson lines in the $\SYMNNk$ representation) and they have the scaling dimension $\Delta=1/2$. Therefore the unitarity bound is saturated and we have four (complex) free bosonic fields\cite{Aharony:2008ug}. The $U(1)\times U(1)$ theory with level $k$ is also a free theory with four free bosonic fields. In the $U(N)\times U(N)$ $k=1$ theory, we could have another interacting part which decouples from the free fields $C_I$.
Without knowing the existence of the interacting part, we can only say that, by exchanging $k$ and $N$, some quantities of the two theories are related, but it needs more investigation to extend this to the full theory.
Nevertheless, it is still interesting to see to what extent this correspondence can be checked.
In the following, we will consider two classes of operators that are related by the exchange of $k$ and $N$.

We have one class of operators of the schematic form $C^k$ and another class of the form $\det(C)$. $C^k$ has $k$ units of $U(1)_b$ charge and $\det(C)$ has $N$ units of $U(1)_b$ charge.
Note that the operator $\det (C)$ has symmetric $SU(4)_R$ indices. Since $C^k$ is in the $k$'th symmetric product of $\mathbf{4}$'s, it is natural for $\det(C)$ to be in the $N$'th symmetric product of $\mathbf{4}$'s.
In Appendix \ref{A:SUGRA}, we confirm explicitly that the operators $C^k$ and $\det (C)$ are in the $k$'th and $N$'th symmetric product of the $\mathbf{4}$ representation, respectively, in the supergravity side by quantizing the collective coordinates.

By $AdS/CFT$ duality \cite{Maldacena:1997re,Witten:1998qj,Gubser:1998bc}, the operator correspondence can also be seen in the string theory side.
The effective field theory on $AdS_4$ in type IIA is given by the following set of equations of motion\cite{Aharony:2008ug}:
\begin{equation}
\begin{split}
&dF^{D0}=0, \qquad dF^{D4}=0\\
&*_4 d *_4 F^{J} = 0\;,\qquad *_4 d *_4 F^K = (N^2 + k^2) e^{2\phi} (da+ A^K)\;.
\end{split}
\end{equation}
Here $F^{D0}$ comes from the two form field $F_2$ and $F^{D4}$ comes from dualizing the four-form $\tilde{F_4}=dA_3-A_1 \wedge H_3$ integrated over the 2-cycle $\CP$, and $A^J$ and $A^K$ are defined by
\begin{equation}
\begin{split}
A^J&=k A^{D0} + N A^{D4}\\
A^K&=N A^{D0} - k A^{D4}\;.
\end{split}
\end{equation}
The gauge field $A^K$ obtains mass due to the axion, but $A^J$ remains massless and the current associated with $A^J$ is invariant under the simultaneous exchange of $k$ and $N$, and $A^{D0}$ and $A^{D4}$.
This is consistent with the previous statement since D0-branes are related to the operators of the form $C^k$ and D4-branes are related to the operators of the form $\det(C)$ \cite{Aharony:2008ug}.
In Appendix \ref{A:SUGRA}, we compute the scaling dimensions of the operators $C^k$ and $\det(C)$ in the supergravity side and check that the result accords with the field theory expectation.
Note that the exchange of $N$ and $k$ means exchanging the values of the six-form flux $*F_4$ on $\CPPP$ and the two-form flux $F_2$ on $\CP$ in $\CPPP$, which are $N$ and $k$ respectively.

D0- and D4-branes in $\CPPP$ have an interesting relation.
Note that a line is dual to a hyperplane in $\Bbb{C}^4$.
Given the standard metric of $\Bbb{C}^4$, then a $\CPP$ has a one-to-one correspondence to a point in $\CPPP$.
Therefore we can relate a D0-brane and a D4-brane in $\CPPP$ in some sense.
Keeping this in mind, let us suppose that there is an associated string duality that exchanges the space $\CPPP$, which consists of lines in $\mathbb{C}^4$, with its dual $\CPPP$, which consists of hyperplanes in $\mathbb{C}^4$.
Then $k$ and $N$ will be exchanged under duality since, as mentioned above, $k$ is the amount of the two form flux $F_2$ and $N$ is the amount of the six-form flux $*F_4$, under which D0-branes and D4-branes are charged, respectively.
We can also argue that the radius of the eleventh direction $R/k$ changes to $R/N$ under duality, where $R$ is the radius of curvature.
Suppose we have a D4-brane wrapped on $\CPP$ in $\CPPP$.
Using the relation between a point and a $\CPP$ in $\CPPP$ mentioned above, we will regard it as a D0-brane.
Then the tension of the new D0-brane $T_{D0 \mathrm{new}}$ will be $T_{D4} \cdot \mathrm{Vol}(\CPP)$, where $T_{D4}$ is the D4-brane tension.
The tension of a D0-brane tells us what the radius of the eleventh direction is.
That is, the eleventh direction in this new setting has the radius $R_{11 \mathrm{new}}=1/T_{D0\mathrm{new}}=(T_{D4} \cdot \mathrm{Vol}(\CPP))^{-1}$.
In Appendix \ref{A:SUGRA}, we calculate $T_{D4} \cdot \mathrm{Vol}(\CPP)$.
Borrowing the result\footnote{Note that the $\CPPP$ radius $R$ is assumed to be the same for both the original and the dual pictures, which also means the Planck length $l_p$ remains invariant.},
\begin{equation}
R_{11 \mathrm{new}}= \frac R N \;.
\end{equation}
That is, the radius of the eleventh direction effectively changes from $R/k$ to $R/N$.

In this section, we started with the assumption that non-integral diagonal monopole configurations are allowed when we perform the path integral in the conformal phase.
Then we speculated about the level-rank duality and provided two clues for this.
One is the supersymmetry enhancement to $\mathcal{N}=8$ when the level $k$ is 1 or 2, or the rank $N$ is 1 or 2.
The other is the relation between the two operators of the form $C^k$ and $\det(C)$.
Since the 't Hooft coupling $\lambda=N/k$ becomes $1/\lambda$, we might be able to see the relation between strong and weak coupling field theories, or gravity theories.
However, as mentioned above, we assumed the existence of non-integral diagonal monopole configurations so this theory is different from the theory with only integral monopoles.
Also, it is not clear at this point whether the correspondence extends to the full theory and it requires further work to clarify this.
Nevertheless, it will still be interesting to find additional properties of the theory that are related by the exchange of the level $k$ and rank $N$.

\section*{Acknowledgments}
It is a pleasure to thank T. Dimofte, A. Kapustin, I. Samsonov, S. Sch\"afer-Nameki, J. Schwarz, J. Song, K. Vyas and M. Yamazaki for helpful discussions.
Especially, I would like to thank H. Ooguri for reading the manuscript carefully and giving valuable advice.
I also thank the students and organizers of International School of Subnuclear Physics at Erice, where part of the work was completed.
This work is supported in part by Samsung Scholarship and DOE grant DE-FG03-92-ER40701.

\appendix

\section*{Appendix}

\section{Dimensions and $SU(4)_R$ representations of the operators $\det(C)$ and $C^k$ in the supergravity approximation}\label{A:SUGRA}
As mentioned in Section 4, in the type IIA description, the operator $\det(C)$ corresponds to the D4-brane wrapped on $\CPP$.
Baryons in $AdS_5\times S^5$ are constructed in \cite{Witten:1998xy,Gross:1998gk} and baryon-like operators similar to this operator are given in \cite{Gubser:1998fp}.
Here, we will calculate the mass of this wrapped D4-brane in the supergravity approximation and relate it to the scaling dimension of the corresponding operator in the field theory following \cite{Witten:1998xy,Gubser:1998fp}.
The metric we consider in type IIA-theory is
\begin{equation}\label{E:fullmetricIIA}
ds^2 = \frac {R^2} 4 ds^2_{AdS_4} + R^2 ds^2_{\CPPP}\;.
\end{equation}
The $AdS_4$ metric is given by
\begin{equation}
ds_{AdS_4}^2=\frac{-dt^2+dx^2+dy^2+dz^2}{z^2}\;,
\end{equation}
whose Ricci tensor satisfies $(R_{AdS_4})_{\mu\nu}=-3(g_{AdS_4})_{\mu\nu}$.
The $\CPPP$ metric is the Fubini-Study metric
\begin{equation}\label{E:metricCP3}
ds^2_{\CPPP}=\frac{\sum_i dz^i d\bar{z}^i}{1+\sum_j z^j \bar{z}^j}-\frac{\sum_i \bar{z}^i dz^i \sum_j z^j d\bar{z}^j}{(1+\sum_k z^k \bar{z}^k)^2}\;,
\end{equation}
where the summation runs from $1$ to $3$ and whose Ricci tensor satisfies $(R_{\CPPP})_{m n}= 8 (g_{\CPPP})_{m n}$.
The mass of the D4-brane wrapped over $\CPP$ is given by
\begin{equation}
m=T_{D4} \cdot \mathrm{Vol}(\CPP)\;,
\end{equation}
where $T_{D4}$ is the tension of the D4-brane:
\begin{equation}
T_{D4}=((2\pi)^4 g_s l_s^5)^{-1}\;.
\end{equation}
We think of $\CPP$ as a three dimensional plane through the origin in $\Bbb{C}^4$, which becomes $\CPPP$ after projective identification.
That is, a typical $\CPP$ in the coordinate system for \eqref{E:metricCP3} is the hyperplane $z^3=0$.
Then $\mathrm{Vol}(\CPP)=\pi^2/2$.
The scaling dimension $\Delta$ for the corresponding operator is given by $\Delta=m R/2$ when $m R$ is large.
The factor $1/2$ is needed to account for the factor $1/4$ in front of the $AdS_4$ part of \eqref{E:fullmetricIIA}.
The radius of the eleventh dimension is $R/k$, which becomes $g_s l_s$ in type IIA-theory.
Also, $l_p^3 = g_s l_s^3$.
The radius $R$ is given by\cite{Aharony:2008ug}
\begin{equation}
R=(2^5 \pi^2 N k)^{1/6} l_p\;.
\end{equation}
Utilizing all these relations, we obtain $\Delta= N/2$ up to corrections that are smaller by a factor of $1/N$.
This is the scaling dimension of $\det(C)$ in the field theory side.

In the same way, we can calculate the scaling dimension of the operator $C^k$.
In the gravity side, this is a D0-brane which is a point in $\CPPP$.
The volume of the D0-brane is 1.
Then, by following the same procedure, we have
\begin{equation}
\Delta= m \frac R 2 = V_{D0} T_{D0} \frac R 2 = \frac 1 {g_s l_s} \frac R 2 = \frac k 2\;,
\end{equation}
which agrees with the field theory anticipation.

We can also quantize the collective coordinates of a D0- or D4-brane to see in which representation of $SU(4)_R$ each of them is.
We closely follow \cite{Witten:1998xy}.
Let us consider a D4-brane first.
Since it is a hyperplane in $\Bbb{C}^4$, we want to consider the quantum wave state on the homogeneous space $G/H$ where $G=SU(4)$ and $H=S(U(3)\times U(1))$.
The state is not an ordinary function on $G/H$, but a section of a line bundle of degree $N$ because the D4-brane is charged under the six-form field $*F_4$ and $N$ units of $*F_4$ flux penetrate $\CPPP$.
A section of a line bundle on $G/H$ of degree $N$ is a function on the $G$ manifold $\psi: G\rightarrow \Bbb{C}$ obeying
\begin{equation}
\psi(gh)=\psi(g)r(h)\;,
\end{equation}
for $g\in G$ and $h\in H$.
$h\mapsto r(h)$ is a homomorphism of $H$ to $U(1)$ and it is given by the product of the trivial homomorphism $x \in SU(3)\mapsto 1\in U(1)$ and the degree $N$ homomorphism from $U(1)$ generated by $\mathrm{diag}(-1,\frac 1 3,\frac 1 3,\frac 1 3)$ to $U(1)$.
Therefore the section $\psi$ is $SU(3)$ invariant and transforms with charge $N$ under $U(1)$.

Now let us consider a unitary $4\times 4$ matrix with elements $g^i_{\;\; j}$, $i,j=1,\cdots, 4$.
$g^i_{\;\; j}$ transform as $(\mathbf{4},\mathbf{\bar{4}})$ under $SU(4)\times SU(4)$, and as $\mathbf{(4,1)^1}\oplus\mathbf{(4,\bar{3})^{-\frac 1 3}}$ under $SU(4)\times S(U(3)\times U(1))$.
The superscript is the $U(1)$ charge.
To make a section of the line bundle of degree $N$, we choose a polynomial of degree $N$ in the $(\mathbf{4,1})^\mathbf{1}$.
This is the lowest degree polynomial that has charge $N$, which minimizes the energy.
Therefore, the D4-brane wrapped on $\CPP$ transforms in the $N$'th symmetric product of $\mathbf{4}$'s of $SU(4)_R$.

In the same way, we see that a D0-brane transforms in the $k$'th symmetric product of $\mathbf{4}$'s of $SU(4)_R$ since it is charged under the two form $F_2$, which has the value $F_2=kJ$ on $\CPPP$ and this determines a line bundle of degree $k$ over $\CPPP$.


%
%


\begin{thebibliography}{1}



\bibitem{Aharony:2008ug}
  O.~Aharony, O.~Bergman, D.~L.~Jafferis and J.~Maldacena,
  ``N=6 superconformal Chern-Simons-matter theories, M2-branes and their
  gravity duals,''
  arXiv:0806.1218 [hep-th].



\bibitem{Schwarz:2004yj}
  J.~H.~Schwarz,
  ``Superconformal Chern-Simons theories,''
  JHEP {\bf 0411}, 078 (2004)
  [arXiv:hep-th/0411077].




\bibitem{Bagger:2006sk}
  J.~Bagger and N.~Lambert,
  ``Modeling multiple M2's,''
  Phys.\ Rev.\  D {\bf 75}, 045020 (2007)
  [arXiv:hep-th/0611108].

\bibitem{Bagger:2007jr}
  J.~Bagger and N.~Lambert,
  ``Gauge Symmetry and Supersymmetry of Multiple M2-Branes,''
  Phys.\ Rev.\  D {\bf 77}, 065008 (2008)
  [arXiv:0711.0955 [hep-th]].

\bibitem{Bagger:2007vi}
  J.~Bagger and N.~Lambert,
  ``Comments On Multiple M2-branes,''
  JHEP {\bf 0802}, 105 (2008)
  [arXiv:0712.3738 [hep-th]].

\bibitem{Gustavsson:2007vu}
  A.~Gustavsson,
  ``Algebraic structures on parallel M2-branes,''
  arXiv:0709.1260 [hep-th].



\bibitem{Krishnan:2008zm}
  C.~Krishnan and C.~Maccaferri,
  ``Membranes on Calibrations,''
  JHEP {\bf 0807}, 005 (2008)
  [arXiv:0805.3125 [hep-th]].





\bibitem{Hosomichi:2008jd}
  K.~Hosomichi, K.~M.~Lee, S.~Lee, S.~Lee and J.~Park,
  ``N=4 Superconformal Chern-Simons Theories with Hyper and Twisted Hyper
  JHEP {\bf 0807}, 091 (2008)
  [arXiv:0805.3662 [hep-th]].





\bibitem{Bandres:2008vf}
  M.~A.~Bandres, A.~E.~Lipstein and J.~H.~Schwarz,
  ``N = 8 Superconformal Chern--Simons Theories,''
  JHEP {\bf 0805}, 025 (2008)
  [arXiv:0803.3242 [hep-th]].

\bibitem{Papadopoulos:2008sk}
  G.~Papadopoulos,
  ``M2-branes, 3-Lie Algebras and Plucker relations,''
  JHEP {\bf 0805}, 054 (2008)
  [arXiv:0804.2662 [hep-th]].


\bibitem{Gauntlett:2008uf}
  J.~P.~Gauntlett and J.~B.~Gutowski,
  ``Constraining Maximally Supersymmetric Membrane Actions,''
  arXiv:0804.3078 [hep-th].



\bibitem{Bandres:2008kj}
  M.~A.~Bandres, A.~E.~Lipstein and J.~H.~Schwarz,
  ``Ghost-Free Superconformal Action for Multiple M2-Branes,''
  arXiv:0806.0054 [hep-th].


\bibitem{Gomis:2008be}
  J.~Gomis, D.~Rodriguez-Gomez, M.~Van Raamsdonk and H.~Verlinde,
  ``Supersymmetric Yang-Mills Theory From Lorentzian Three-Algebras,''
  arXiv:0806.0738 [hep-th].








\bibitem{Benvenuti:2008bt}
  S.~Benvenuti, D.~Rodriguez-Gomez, E.~Tonni and H.~Verlinde,
  ``N=8 superconformal gauge theories and M2 branes,''
  arXiv:0805.1087 [hep-th].









\bibitem{Benna:2008zy}
  M.~Benna, I.~Klebanov, T.~Klose and M.~Smedback,
  ``Superconformal Chern-Simons Theories and $AdS_4/CFT_3$ Correspondence,''
  arXiv:0806.1519 [hep-th].

\bibitem{Bhattacharya:2008bja}
  J.~Bhattacharya and S.~Minwalla,
  ``Superconformal Indices for ${\mathcal N}=6$ Chern Simons Theories,''
  arXiv:0806.3251 [hep-th].

\bibitem{Nishioka:2008gz}
  T.~Nishioka and T.~Takayanagi,
  ``On Type IIA Penrose Limit and N=6 Chern-Simons Theories,''
  arXiv:0806.3391 [hep-th].


\bibitem{Minahan:2008hf}
  J.~A.~Minahan and K.~Zarembo,
  ``The Bethe ansatz for superconformal Chern-Simons,''
  arXiv:0806.3951 [hep-th].

\bibitem{Gaiotto:2008cg}
  D.~Gaiotto, S.~Giombi and X.~Yin,
  ``Spin Chains in N=6 Superconformal Chern-Simons-Matter Theory,''
  arXiv:0806.4589 [hep-th].



\bibitem{Ahn:2008gda}
  C.~Ahn,
  ``Towards Holographic Gravity Dual of N=1Superconformal Chern-Simons Gauge
  Theory,''
  JHEP {\bf 0807}, 101 (2008)
  [arXiv:0806.4807 [hep-th]].




\bibitem{Grignani:2008is}
  G.~Grignani, T.~Harmark and M.~Orselli,
  ``The SU(2) x SU(2) sector in the string dual of N=6 superconformal
  Chern-Simons theory,''
  arXiv:0806.4959 [hep-th].





\bibitem{Hosomichi:2008jb}
  K.~Hosomichi, K.~M.~Lee, S.~Lee, S.~Lee and J.~Park,
  ``N=5,6 Superconformal Chern-Simons Theories and M2-branes on Orbifolds,''
  arXiv:0806.4977 [hep-th].

\bibitem{Bagger:2008se}
  J.~Bagger and N.~Lambert,
  ``Three-Algebras and N=6 Chern-Simons Gauge Theories,''
  arXiv:0807.0163 [hep-th].


\bibitem{Bak:2008cp}
  D.~Bak and S.~J.~Rey,
  ``Integrable Spin Chain in Superconformal Chern-Simons Theory,''
  JHEP {\bf 0810}, 053 (2008)
  [arXiv:0807.2063 [hep-th]].



\bibitem{McLoughlin:2008ms}
  T.~McLoughlin and R.~Roiban,
  ``Spinning strings at one-loop in $AdS_4 \times P^3$,''
  arXiv:0807.3965 [hep-th].


\bibitem{Alday:2008ut}
  L.~F.~Alday, G.~Arutyunov and D.~Bykov,
  ``Semiclassical Quantization of Spinning Strings in $AdS_4 \times CP^3$,''
  arXiv:0807.4400 [hep-th].




\bibitem{Krishnan:2008zs}
  C.~Krishnan,
  ``AdS4/CFT3 at One Loop,''
  JHEP {\bf 0809}, 092 (2008)
  [arXiv:0807.4561 [hep-th]].



\bibitem{Ooguri:2008dk}
  H.~Ooguri and C.~S.~Park,
  ``Superconformal Chern-Simons Theories and the Squashed Seven Sphere,''
  arXiv:0808.0500 [hep-th].




\bibitem{Jafferis:2008qz}
  D.~L.~Jafferis and A.~Tomasiello,
  ``A simple class of N=3 gauge/gravity duals,''
  arXiv:0808.0864 [hep-th].


\bibitem{Berenstein:2008dc}
  D.~Berenstein and D.~Trancanelli,
  ``Three-dimensional N=6 SCFT's and their membrane dynamics,''
  arXiv:0808.2503 [hep-th].


\bibitem{Hosomichi:2008ip}
  K.~Hosomichi, K.~M.~Lee, S.~Lee, S.~Lee, J.~Park and P.~Yi,
  ``A Nonperturbative Test of M2-Brane Theory,''
  arXiv:0809.1771 [hep-th].


\bibitem{Ahn:2008ua}
  C.~Ahn,
  ``Squashing Gravity Dual of N=6 Superconformal Chern-Simons Gauge Theory,''
  arXiv:0809.3684 [hep-th].

\bibitem{Klebanov:2008vq}
  I.~Klebanov, T.~Klose and A.~Murugan,
  ``$AdS_4/CFT_3$ -- Squashed, Stretched and Warped,''
  arXiv:0809.3773 [hep-th].


\bibitem{McLoughlin:2008he}
  T.~McLoughlin, R.~Roiban and A.~A.~Tseytlin,
  ``Quantum spinning strings in $AdS_4 \times CP^3$: testing the Bethe Ansatz
  proposal,''
  arXiv:0809.4038 [hep-th].












\bibitem{Moore:1989yh}
  G.~W.~Moore and N.~Seiberg,
  ``Taming the Conformal Zoo,''
  Phys.\ Lett.\  B {\bf 220}, 422 (1989).


\bibitem{Itzhaki:2002rc}
  N.~Itzhaki,
  ``Anyons, 't Hooft loops and a generalized connection in three dimensions,''
  Phys.\ Rev.\  D {\bf 67}, 065008 (2003)
  [arXiv:hep-th/0211140].



\bibitem{'tHooft:1977hy}
  G.~'t Hooft,
  ``On The Phase Transition Towards Permanent Quark Confinement,''
  Nucl.\ Phys.\  B {\bf 138}, 1 (1978).


\bibitem{Borokhov:2002ib}
  V.~Borokhov, A.~Kapustin and X.~k.~Wu,
  ``Topological disorder operators in three-dimensional conformal field
  theory,''
  JHEP {\bf 0211}, 049 (2002)
  [arXiv:hep-th/0206054].




\bibitem{Borokhov:2002cg}
  V.~Borokhov, A.~Kapustin and X.~k.~Wu,
  ``Monopole operators and mirror symmetry in three dimensions,''
  JHEP {\bf 0212}, 044 (2002)
  [arXiv:hep-th/0207074].



\bibitem{Kapustin:2005py}
  A.~Kapustin,
  ``Wilson-'t Hooft operators in four-dimensional gauge theories and
  S-duality,''
  Phys.\ Rev.\  D {\bf 74}, 025005 (2006)
  [arXiv:hep-th/0501015].



\bibitem{Lambert:2008et}
  N.~Lambert and D.~Tong,
  ``Membranes on an Orbifold,''
  Phys.\ Rev.\ Lett.\  {\bf 101}, 041602 (2008)
  [arXiv:0804.1114 [hep-th]].




\bibitem{Distler:2008mk}
  J.~Distler, S.~Mukhi, C.~Papageorgakis and M.~Van Raamsdonk,
  ``M2-branes on M-folds,''
  JHEP {\bf 0805}, 038 (2008)
  [arXiv:0804.1256 [hep-th]].







\bibitem{Klebanov:1998hh}
  I.~R.~Klebanov and E.~Witten,
  ``Superconformal field theory on threebranes at a Calabi-Yau  singularity,''
  Nucl.\ Phys.\  B {\bf 536}, 199 (1998)
  [arXiv:hep-th/9807080].


\bibitem{Gubser:1998fp}
  S.~S.~Gubser and I.~R.~Klebanov,
  ``Baryons and domain walls in an N = 1 superconformal gauge theory,''
  Phys.\ Rev.\  D {\bf 58}, 125025 (1998)
  [arXiv:hep-th/9808075].









\bibitem{Bandres:2008ry}
  M.~A.~Bandres, A.~E.~Lipstein and J.~H.~Schwarz,
  ``Studies of the ABJM Theory in a Formulation with Manifest SU(4)
  R-Symmetry,''
  arXiv:0807.0880 [hep-th].











\bibitem{Naculich:1990hg}
  S.~G.~Naculich and H.~J.~Schnitzer,
  ``Duality Between $SU(N)_K$ And $SU(K)_N$ WZW Models,''
  Nucl.\ Phys.\  B {\bf 347}, 687 (1990).




\bibitem{Maldacena:1997re}
  J.~M.~Maldacena,
  ``The large N limit of superconformal field theories and supergravity,''
  Adv.\ Theor.\ Math.\ Phys.\  {\bf 2}, 231 (1998)
  [Int.\ J.\ Theor.\ Phys.\  {\bf 38}, 1113 (1999)]
  [arXiv:hep-th/9711200].


\bibitem{Witten:1998qj}
  E.~Witten,
  ``Anti-de Sitter space and holography,''
  Adv.\ Theor.\ Math.\ Phys.\  {\bf 2}, 253 (1998)
  [arXiv:hep-th/9802150].


\bibitem{Gubser:1998bc}
  S.~S.~Gubser, I.~R.~Klebanov and A.~M.~Polyakov,
  ``Gauge theory correlators from non-critical string theory,''
  Phys.\ Lett.\  B {\bf 428}, 105 (1998)
  [arXiv:hep-th/9802109].



\bibitem{Gubser:2002tv}
  S.~S.~Gubser, I.~R.~Klebanov and A.~M.~Polyakov,
  ``A semi-classical limit of the gauge/string correspondence,''
  Nucl.\ Phys.\  B {\bf 636}, 99 (2002)
  [arXiv:hep-th/0204051].










\bibitem{Witten:1998xy}
  E.~Witten,
  ``Baryons and branes in anti de Sitter space,''
  JHEP {\bf 9807}, 006 (1998)
  [arXiv:hep-th/9805112].

\bibitem{Gross:1998gk}
  D.~J.~Gross and H.~Ooguri,
  ``Aspects of large N gauge theory dynamics as seen by string theory,''
  Phys.\ Rev.\  D {\bf 58}, 106002 (1998)
  [arXiv:hep-th/9805129].








\end{thebibliography}
\end{document}